\documentstyle[11pt]{article}
\begin{document}
\hfuzz=5pt
\chardef\ii="10
\def\e{\hbox{e}}
\def\i{\hbox{i}}
\def\ih{{\i\over\hbar}}
\def\CD{{\cal D}}
\def\CH{{\cal H}}
\def\CL{{\cal L}}
\def\half{{1\over2}}
\def\bhalf{\hbox{$\half$}}
\def\viert{{1\over4}}
\def\SU{\hbox{SU}}
\def\SO{\hbox{SO}}
\def\O{\hbox{O}}
\def\diag{\hbox{diag}}
\def\erf{\hbox{erf}}
\def\erfc{\hbox{erfc}}
\def\det{\hbox{det}}
\def\bbbr{{\rm I\!R}} 
\def\bbbn{{\rm I\!N}} 
\def\bbbc{{\mathchoice {\setbox0=\hbox{$\displaystyle\rm C$}\hbox{\hbox
to0pt{\kern0.4\wd0\vrule height0.9\ht0\hss}\box0}}
{\setbox0=\hbox{$\textstyle\rm C$}\hbox{\hbox
to0pt{\kern0.4\wd0\vrule height0.9\ht0\hss}\box0}}
{\setbox0=\hbox{$\scriptstyle\rm C$}\hbox{\hbox
to0pt{\kern0.4\wd0\vrule height0.9\ht0\hss}\box0}}
{\setbox0=\hbox{$\scriptscriptstyle\rm C$}\hbox{\hbox
to0pt{\kern0.4\wd0\vrule height0.9\ht0\hss}\box0}}}}
\def\bbbz{{\mathchoice {\hbox{$\sf\textstyle Z\kern-0.4em Z$}}
{\hbox{$\sf\textstyle Z\kern-0.4em Z$}}
{\hbox{$\sf\scriptstyle Z\kern-0.3em Z$}}
{\hbox{$\sf\scriptscriptstyle Z\kern-0.2em Z$}}}}
\large
\begin{titlepage}
\centerline{\normalsize DESY 93 - 140 \hfill ISSN 0418 - 9833}
\centerline{\hfill hep-th/9310162}
\centerline{\normalsize October 1993\hfill}
\vskip.3in
\begin{center}
{\Large ON THE PATH INTEGRAL IN IMAGINARY LOBACHEVSKY SPACE}
\vskip.3in
{\Large Christian Grosche$^*$}
\vskip.3in
{\normalsize\em II.\ Institut f\"ur Theoretische Physik}
\vskip.05in
{\normalsize\em Universit\"at Hamburg, Luruper Chaussee 149}
\vskip.05in
{\normalsize\em 22761 Hamburg, Germany}
\end{center}
\normalsize
\vfill
\begin{center}
{ABSTRACT}
\end{center}
\smallskip
\noindent
The path integral on the single-sheeted hyperboloid, i.e.\ in
$D$-dimensional imaginary Lobachevsky space, is evaluated. A potential
problem which we call ``Kepler-problem'', and the case of a constant
magnetic field are also discussed.

\bigskip\noindent
\centerline{\vrule height0.25pt depth0.25pt width4cm\hfill}
\noindent
{\footnotesize $^*$ Supported by Deutsche Forschungsgemeinschaft under
        contract number GR 1031/2--1.}
\end{titlepage}

 \begin{titlepage}
 \begin{center}
 \ \ \
 \end{center}
 \end{titlepage}
%

\normalsize
\noindent{\bf 1.\ Introduction.}
\vglue 0.4truecm
\noindent
Motion on spaces with constant curvature, positive as well as negative
constant curvature, is of particular interest and appears in several
topics in theoretical physics. Let us e.g.\ mention string theory where
the perturbative expansion \`a la Polyakov \cite{GSW} leads to the
consideration of determinants of Laplacians on Riemann surfaces of
arbitrary genus, a theory where the underlying space is the Poincar\'e,
respectively Lobachevsky space, a space of constant negative curvature.

Another example is the Kepler problem in spaces of constant curvature
\cite{BIJ, GROg}. Here one is interested in the comparison of the
symmetry properties of this problem, where one e.g.\ finds that the
coordinate systems which separate the Kepler problem in spaces of
constant curvature are only two, namely the (pseudo-) spherical and the
(pseudo-) conical, whereas in flat space there are four \cite{LL}.

The evaluations of propagators and its short-time behaviour, and Green
functions are also important in cosmological models. They appear in
several models derived from the Wheeler-DeWitt equation and quantum
gravity, respectively, and lead in a natural way to models, respectively
spaces, with constant curvature. As the simplest case one can study
the free motion in these spaces. Here several models can appear in the
case of constant negative curvature: The single-sheeted and the
two-sheeted hyperboloid. Most simply, they are studied in the
two-dimensional case. The two-sheeted hyperboloid is a particular
realization of the Poincar\'e plane, where only one sheet has been
taken, whereas the single-sheeted has different properties and has not
been studied in such great detail as the former one. However, some
contributions exist, mainly by Gel'fand, Graev and Vilenkin who call
this space imaginary Lobachevsky space \cite{GEGR, GGV, VIL} and have
studied its geometrical structure and group theoretical properties. It
has the peculiarity that the distance $r$ of two points defined by
$\cosh kr$ ($k$ is the curvature) may be positive {\it and\/} imaginary
because $\cosh kr\in[0, \infty)$, i.e.\ it is a space-like set, in
comparison to the usual two-sheeted hyperboloid, also called
``pseudosphere'', which is a time-like set. From the point of view of
special relativity, Lobachevskian models are of interest because the
velocity space, say, possess a constant negative curvature (equal to
$1/c^2$), and the single-sheeted hyperboloid in particular corresponds
to the unphysical region of the variables.

In this note I want to study the path integral on the (D-1)-dimensional
single-sheeted hyperboloid, denoted in the following by $\CH_{-1}^{(D)}$
which is done in Section 2. This special system was not subject to path
integration until now. The path integral on the pseudosphere has been
intensively studied in Refs.~\cite{BJb, GROg, GROc, GRSa}, with its
higher dimensional generalizations in \cite{GROn, GRSc}. In comparison
to this usual pseudosphere we will find that in the case of $\CH_{-1}^{(
D)}$ bound states can appear, depending on the angular momentum number,
which is not possible for the quantum motion on the pseudosphere.

As we will see in Section 3, also a potential problem on the
single-sheeted hyperboloid can be discussed, which will be called
``Kepler-problem'' on the single-sheeted hyperboloid.

In Section 4, the case of a constant magnetic field on $\CH_{-1}^{(D)}$
will be discussed, Section 5 contains some concluding remarks, and in
the Appendix the path integral identity for the modified
P\"oschl-Teller potential is given.
\goodbreak


\vglue 0.6truecm
\noindent{\bf 2.\ The Path Integral.}
\vglue 0.4truecm
\noindent
For simplicity we first consider the simplest case, i.e.~$D=3$. We
start with the equation for the two-dimensional single-sheeted
hyperboloid $\CH_{-1}^{(3)}$ $(1/k=R>0)$
\begin{equation}
   (\vec x,\vec x)=x_0^2-x_1^2-x_2^2=-R^2\enspace.
\end{equation}
We introduce pseudo-spherical polar coordinates on $\CH_{-1}^{(3)}$
\begin{equation}
  x_0=R\sinh\tau\enspace,\qquad
  x_1=R\cosh\tau\sin\phi\enspace,\qquad
  x_2=R\cosh\tau\cos\phi\enspace,
\end{equation}
where $\tau\in(-\infty,\infty)$ and $\phi\in[0,2\pi)$. The addition
theorem on the single-sheeted hyperboloid $\CH_{-1}^{(3)}$ has the form
$(\vec x,\vec y\in\CH_{-1}^{(3)})$
\begin{eqnarray}
  \cosh kr& =&{\vec x_1\cdot\vec x_2\over |\vec x_1|\cdot|\vec x_2|}
  \\
  &=&\Big(\sinh\tau_1\sinh\tau_2-\cosh\tau_1\cosh\tau_2
  \cos(\phi_2-\phi_1)\Big)\enspace.
\end{eqnarray}
We find for the metric: $(g_{ab})=R^2\diag(1,-\cosh^2\tau)$, and
therefore $g=\sqrt{|\det(g_{ab})|}=R^2\cosh\tau$. According to the
canonical formalism \cite{DEW, FEY, GROa, GRSb, GRSg} we construct
the path integral on $\CH_{-1}^{(3)}$ as follows ($T=t''-t'$)
\begin{eqnarray}
  & &
  \!\!\!\!\!\!\!\!\!\!\!\!
  K^{(\CH_{-1}^{(3)})}(x_1'',x_1',x_2'',x_2',x_3'',x_3';T)
  \equiv
  K^{(\CH_{-1}^{(3)})}(\tau'',\tau',\phi'',\phi';T)
  \nonumber\\   & &
  =\int\limits_{x_1(t')=x_1'}^{x_1(t'')=x_1''}\CD x_1(t)
   \int\limits_{x_2t')=x_2'}^{x_2(t'')=x_2''}\CD x_2(t)
   \int\limits_{x_3(t')=x_3'}^{x_3(t'')=x_3''}\CD x_3(t)
   \exp\left[{\i m\over2\hbar}\int_{t'}^{t''}\Big(
   \dot x_1^2-\dot x_2^2-\dot x_3^2\Big)dt\right]
  \\            & &
  \to{1\over R^2}
   \lim_{N\to\infty}\bigg({mR^2\over 2\pi\epsilon\hbar}\bigg)^{N}
  \prod_{j=1}^{N-1}\int_{-\infty}^\infty\cosh\tau_jd\tau_j
  \int_0^{2\pi}d\phi_j
  \nonumber\\   & &\qquad\times
  \exp\left\{\ih\sum_{j=1}^{N}\left[{mR^2\over2\epsilon}\Big(
   \Delta^2\tau_j-{\widehat{\cosh}}^2\tau_j\Delta^2\phi_j\Big)
 -{\epsilon\hbar^2\over8mR^2}\bigg(1+{1\over\cosh^2\tau_j}\bigg)
   \right]\right\}
  \\            & &
   \equiv{1\over R^2}
   \int\limits_{\tau(t')=\tau'}^{\tau(t'')=\tau''}\cosh\tau\CD\tau(t)
   \int\limits_{\phi(t'')=\phi''}^{\phi(t'')=\phi''}\CD\phi(t)
  \nonumber\\   & &\qquad\times
   \exp\left\{\ih\int_{t'}^{t''}\left[{m\over2}R^2\Big(
   \dot\tau^2-\cosh^2\tau\dot\phi^2\Big)
   -{\hbar^2\over8mR^2}\bigg(1+{1\over\cosh^2\tau}\bigg)
  \right]dt\right\}\enspace.
\end{eqnarray}
Here are $\epsilon=T/N$, $q_j=q(t'+j\epsilon)$, ${\widehat f}^2(q_j)=
f(q_j)f(q_{j-1})$ for any function of the coordinates $\tau$ and $\phi$,
and $j=0,\dots,N$, $\Delta q_j=q_j-q_{j-1}$, and we interpret the
limit $N\to\infty$ as equivalent with $\epsilon\to0$, $T$ fixed. Note
that due to the indefinite metric the factors ``$\i$'' in the ``measure
term'' cancel each other \cite{BJb}. The corresponding short-time
propagator is given by
\begin{eqnarray}
  & &K^{(\CH_{-1}^{(3)})}(\tau_j,\tau_{j-1},\phi_j,\phi_{j-1};\epsilon)
  ={m\over 2\pi\epsilon\hbar}\cosh\tau_j
  \nonumber\\   & &\qquad\times
   \exp\left\{\ih\left[{mR^2\over2\epsilon}\Big(
   \Delta^2\tau_j-{\widehat{\cosh}}^2\tau_j\Delta^2\phi_j\Big)
   -{\epsilon\hbar^2\over8mR^2}\bigg(1+{1\over\cosh^2\tau_j}\bigg)
   \right]\right\}\enspace.
\end{eqnarray}
Note that the pre-exponential factor does not depend on $R$. The
$\phi$-path integration can be separated \cite{GROj} immediately and
we obtain
\begin{equation}
   K^{(\CH_{-1}^{(3)})}(\tau'',\tau',\phi'',\phi';T)
   ={\e^{-\i\hbar T/8mR^2}\over\big(-\cosh\tau'\cosh\tau'')^{1/2}}
   \sum_{l=-\infty}^{\infty}{\e^{\i l(\phi''-\phi')}\over2\pi R^2}
   K^{(\CH_{-1}^{(3)})}_l(\tau'',\tau';T)
\end{equation}
with $K^{(\CH_{-1}^{(3)})}_l(\tau'',\tau';T)$ given by
\begin{equation}
  K^{(\CH_{-1}^{(3)})}_l(\tau'',\tau';T)
  =\int\limits_{\tau(t')=\tau'}^{\tau(t'')=\tau''}\CD\tau(t)
  \exp\left\{\ih\int_{t'}^{t''}\left[{m\over2}R^2\dot\tau^2
  +{\hbar^2\over2mR^2}{l^2-1/4\over\cosh^2\tau}\right]dt\right\}
  \enspace,
\end{equation}
which is a usual one-dimensional path integral. This path integral has
the form of the special case of the modified P\"oschl-Teller potential
as sketched in the Appendix. Therefore we can write down the solution
of the path integral on the single-sheeted hyperboloid
($n=0,1,\dots,N_M<|l|-\half$)
\begin{eqnarray}
  & &{1\over R^2}\ih \int_0^{\infty}  dT\,\e^{\i TE/\hbar}
  \int\limits_{\tau(t')=\tau'}^{\tau(t'')=\tau''}\cosh\tau\CD\tau(t)
  \int\limits_{\phi(t')=\phi'}^{\phi(t'')=\phi''}\CD\phi(t)
  \nonumber\\   & &\qquad\times
  \exp\left\{\ih\int_{t'}^{t''}\left[{m\over2}R^2
  (\dot\tau^2-\cosh^2\tau\dot\phi^2)-{\hbar^2\over8mR^2}
  \bigg(1+{1\over\cosh^2\tau}\bigg)\right]dt\right\}
  \nonumber\\   & &\quad
  =\big(-\cosh\tau'\cosh\tau'')^{-1/2}\sum_{l=-\infty}^{\infty}
   {\e^{\i l(\phi''-\phi')}\over2\pi}
  \nonumber\\   & &\qquad\times
  {m\over\hbar^2}
  \Gamma\left(\sqrt{-{2mR^2E\over\hbar^2}+{1\over4}}-|l|+\half\right)
  \Gamma\left(\sqrt{-{2mR^2E\over\hbar^2}+{1\over4}}+|l|+\half\right)
  \nonumber\\   & &\qquad\times
  P_{|l|-1/2}^{-\sqrt{-2mR^2E/\hbar^2+1/4}}(\tanh\tau_<)
  P_{|l|-1/2}^{-\sqrt{-2mR^2E/\hbar^2+1/4}}(-\tanh\tau_>)
           \\   & &\quad
  =\big(-\cosh\tau'\cosh\tau'')^{-1/2}\sum_{l=-\infty}^{\infty}
   {\e^{\i l(\phi''-\phi')}\over2\pi R^2}
  \nonumber\\   & &\qquad\times
  \left[\sum_{n=0}^{N_M}\bigg(n-|l|-\half\bigg){\Gamma(2|l|-n)\over n!}
  \frac{P_{|l|-1/2}^{n-|l|+\half}(\tanh\tau')
  P_{|l|-1/2}^{n-|l|+\half}(\tanh\tau'')}
  {-\hbar^2[(n-|l|+\half)^2-1/4]/2mR^2-E}
  \right.\nonumber\\   & &\qquad\qquad\left.
  +\half\int_{-\infty}^{\infty}\,
  {dp\,p\sinh\pi p\over\hbar^2(p^2+1/4)/2mR^2-E}
   {P^{\i p}_{|l|-1/2}(\tanh\tau'')P^{-\i p}_{|l|-1/2}(\tanh\tau')
   \over\cos^2\pi l+\sinh^2\pi p}\right]\enspace.\qquad
\label{numf}
\end{eqnarray}
Wave-functions and energy spectrum are easily read off from the spectral
expansion (\ref{numf}). Note that for $l\not=0$ there are bound states.
The generalization to higher dimensions can be done in a straightforward
way, by replacing the circular wave-functions by the hyperspherical
harmonics $S_l^\mu(\Omega)$, and the quantum number $l\in\bbbz$ by the
corresponding principle quantum number $l\in\bbbn_0$, including the
appropriate changes in the effective Lagrangian, and with the prefactor
replaced by $R^{1-D}$. In order to to this we introduce the
(pseudo-bispherical) coordinate system \cite{BJb, GEGR, GGV, VIL}
\begin{eqnarray}
 & &x_0=R\sinh\tau\enspace,
  \nonumber\\
 & &x_1=R\cosh\tau\cos\theta_{D-2}\enspace,
  \nonumber\\
 & &x_2=R\cosh\tau\sin\theta_{D-2}\cos\theta_{D-3}\enspace,
  \nonumber\\
 & &\qquad\vdots
  \nonumber\\
 & &x_{D-2}=R\cosh\tau\sin\theta_{D-2}\dots\cos\theta_2\cos\phi\enspace,
  \nonumber\\
 & &x_{D-1}=R\cosh\tau\sin\theta_{D-2}\dots\cos\theta_2\sin\phi\enspace,
\end{eqnarray}
where $\tau\in(-\infty,\infty)$, $\theta_1\equiv\phi\in[0,2\pi)$, and
$\theta_k\in[0,\pi)$, $k=2,\dots,D-2$. The metric tensor on the
(D-1)-dimensional single-sheeted hyperboloid is given by: $(g_{ab})=R^2
\diag(1,-\cosh^2\tau, -\cosh^2\tau$\linebreak$\times\sin^2\theta_{D-2},
\dots, -\cosh^2\tau\dots\sin^2\theta_2)$ $(a,b=1,\dots, D-1)$.
Therefore we obtain for the Hamiltonian on $\CH_{-1}^{(D)}$
\begin{eqnarray}
  H&= &-{\hbar^2\over2mR^2}\Bigg\{
  \Bigg[{\partial^2\over\partial\tau^2}
       +(D-2)\tanh\tau{\partial\over\partial\tau}\Bigg]
  \nonumber\\   & &
  -{1\over\cosh^2\tau}\Bigg[{\partial^2\over\partial\theta^2_{D-2}}
       +(D-3)\coth\theta_{D-2}{\partial\over\partial\theta_{D-2}}\Bigg]-
  \dots-{1\over\cosh^2\tau\dots\sin^2\theta_2}
  {\partial^2\over\partial\phi}\Bigg\}\qquad
           \\   &=&
  {1\over2mR^2}\left[p_\tau^2-{1\over\cosh^2\tau}p^2_{\theta_{D-2}}-
  \dots-{1\over\cosh^2\tau\dots\sin^2\theta_2}p_\phi^2\right]
  +\Delta V(\tau,\{\theta\})\enspace,
\end{eqnarray}
with the quantum potential
\begin{equation}
\Delta V(\tau,\{\theta\})={\hbar^2\over8mR^2}\bigg[(D-2)^2
  +{1\over\cosh^2\tau}+\dots+
  {1\over\cosh^2\tau\dots\sin^2\theta_2}\bigg]
\end{equation}
[$\{\theta\}$ denotes the set of variables $\theta_k$ ($k=1,\dots,
D-2$)]. Furthermore ($g=\det(g_{ab})=\cosh^{D-2}\tau$\linebreak
$\times\prod_{k=1}^{D-2}(\sin\theta_k)^{k-1}$)
\begin{equation}
  p_a={\hbar\over\i}\bigg({\partial\over\partial q^a}
                    +{\Gamma_a\over2}\bigg),
  \qquad \Gamma_a={\partial\log\sqrt{g}\over\partial q^a}\enspace.
\end{equation}
Thus we obtain for the (Lagrangian) path integral on $\CH_{-1}^{(D)}$
[$\vec x=(x_0,\dots,x_{D-1})$]
\begin{eqnarray}
  & &\!\!\!\!\!\!
  K^{(\CH_{-1}^{(D)})}(\vec x'',\vec x';T)
  \equiv
  K^{(\CH_{-1}^{(D)})}(\tau'',\tau,\{\theta''\},\{\theta'\};T)
  \nonumber\\   & &\!\!\!\!\!\!
  =\prod_{k=0}^{D-1}
   \int\limits_{x_k(t')=x_k'}^{x_k(t'')=x_k''}\CD x_k(t)
   \exp\left[{\i m\over2\hbar}\int_{t'}^{t''}\Bigg(
   \dot x_0^2-\sum_{k=1}^{D-1}\dot x_k^2\Bigg)dt\right]
  \nonumber\\   & &\!\!\!\!\!\!
  \to\!\int\limits_{\tau(t')=\tau'}^{\tau(t'')=\tau''}
  \!\cosh^{D-2}\tau\CD\tau(t)\!
   \int\limits_{\Omega(t')=\Omega'}^{\Omega(t'')=\Omega''}\!\CD\Omega(t)
  \exp\Bigg\{\ih\int_{t'}^{t''}\Big[
  \CL_{Cl}(\tau,\dot\tau,\{\theta\},\{\dot\theta\})
  -\Delta V(\tau,\{\theta\})\Big]\Bigg\}
  \nonumber\\   & &\!\!\!\!\!\!
  =R^{1-D}\lim_{N\to\infty}
  \bigg({mR^2\over2\pi\i\epsilon\hbar}\bigg)^{{N\over2}(D-2)}
  \bigg({\i mR^2\over2\pi\hbar\epsilon}\bigg)^{{N\over2}}
  \prod_{j=1}^{N-1}\int_0^\infty\cosh^{D-2}\tau_jd\tau_j
  \int d\Omega_j
  \nonumber\\   & &\qquad\times
  \exp\left\{\ih\sum_{j=1}^N
  \Big[\CL_{Cl}^N(\tau_{j-1},\tau_j,\{\theta_{j-1}\},\{\theta_j\})
  -\epsilon\Delta V(\tau_j,\{\theta_j\})\Big]\right\}\enspace.
\end{eqnarray}
$\CL_{Cl}$ is the classical Lagrangian
\begin{equation}
  \CL_{Cl}(\tau,\dot\tau,\{\theta\},\{\dot\theta\})={mR^2\over2}
  \Big[\dot\tau^2-\cosh^2\tau\,\dot\theta^2_{D-2}-\dots-
  (\cosh^2\tau\dots\sin^2\theta_2)\dot\phi^2\Big]\enspace,
\end{equation}
and its counterpart on the lattice reads
\begin{eqnarray}
  & &\CL_{Cl}^N(\tau_{j-1},\tau_j,\{\theta_{j-1}\},\{\theta_j\})
  \nonumber\\   & &\qquad
  ={mR^2\over2\epsilon^2}\Big[\Delta^2\tau_j
  -{\widehat{\cosh^2\tau_j}}\Delta^2\theta_{D-2,j}-\dots-
  ({\widehat{\cosh^2\tau_j}}\dots
   {\widehat{\sin^2\theta_{2,j}}})\Delta^2\phi_j\Big]\enspace.\qquad
\end{eqnarray}
$d\Omega=\prod_{k=1}^{D-2}(\sin\theta_k)^{k-1}d\theta_k$ is the
$(D-2)$-dimensional surface element on the unit-sphere $S^{(D-2)}$.
Note again that the pre-exponential factor in the short-time kernel
does not depend on $R$.

Due to the very singular nature of $\Delta V(\tau,\{\theta\})$ this
path integral is at it stands not tractable. However, we can use a path
integral identity (based on a method developed in \cite{GJ, PS})
already derived in \cite{GRSb, GRSc} to simplify the path integration
significantly and separate the angular variables $\theta_{D-2}, \dots,
\phi$ from the hyperbolic coordinate $\tau$. I introduce the quantity
$\psi^{(','')}$ defined by
\begin{equation}
  \cos\psi^{(','')}=\cos\theta_{D-2}'\cos\theta_{D-2}''
  +\sum_{m=1}^{D-3}\cos\theta'_m\cos\theta_m''
  \prod_{n=m+1}^{D-2}\sin\theta^{''}_n\sin\theta^{''}_n+
  \prod_{n=1}^{D-2}\sin\theta'_n\sin\theta''_n\enspace,
\end{equation}
which is actually the addition theorem on the $S^{(D-2)}$-sphere and
$\cos\psi^{(','')}=\Omega'\cdot\Omega''$, where $\Omega^{(','')}$
are unit vectors on the $S^{(D-2)}$-sphere. Using the result of
\cite{GRSb} the following path integral identity can be achieved
(replace $R^2=r_j^2=-\cosh^2\tau_j$ in (2.27) in \cite{GRSb})
\begin{eqnarray}
  & &
  \exp\bigg\{-{\i mR^2\over2\epsilon\hbar}{\widehat{\cosh^2\tau_j}}
  \Big[\Delta^2\theta_{D-2,j}+\dots+
  ({\widehat{\sin^2\theta_{D-2,j}}}\dots
   {\widehat{\sin^2\theta_{2,j}}})\Delta^2\phi_j\Big]\bigg\}
  \nonumber\\   & &\dot=
  \exp\bigg[-{\i mR^2\over\epsilon\hbar}\widehat{\cosh^2\tau_j}
  (1-\cos\psi_{j-1,j})
  \nonumber\\   & &\qquad\qquad
  +{\i\epsilon\hbar\over8mR^2\cosh^2\tau_j}\bigg(1+
  {1\over\sin^2\theta_{D-2,j}}+\dots+{1\over\sin^2\theta_{D-2,j}\dots
   \sin^2\theta_{2,j}}\bigg)\bigg]\enspace.
\end{eqnarray}
Here I have used the symbol $\dot=$ - following DeWitt \cite{DEW}- to
denote ``equivalence as far as use in the path integral is concerned''.
The highly singular terms are cancelling and I obtain
\begin{eqnarray}
  & &K^{(\CH_{-1}^{(D)})}(\tau'',\tau',\{\theta''\},\{\theta'\};T)
  =R^{1-D}\int\limits_{\tau(t')=\tau'}^{\tau(t'')=\tau''}
  \cosh^{D-2}\tau\CD\tau(t)
   \int\limits_{\Omega(t')=\Omega'}^{\Omega(t'')=\Omega''}\CD\Omega(t)
  \nonumber\\   & &\qquad\qquad\qquad\qquad\times
  \exp\left\{{\i mR^2\over2\hbar}
      \int_{t'}^{t''}\Big[\dot\tau^2-2\cosh^2\tau(1-\cos\psi)\Big]
  -{\i\hbar T(D-2)^2\over8mR^2}\right\}\enspace.\qquad
\end{eqnarray}
Expanding now the exponential according to \cite{GRA}, p.980:
\begin{equation}
  \e^{z\cos\psi^{(','')}}
  =\bigg({2\over z}\bigg)^\nu\Gamma(\nu)\sum_{l=0}^\infty(l+\nu)
   C_l^\nu(\cos\psi^{(','')}) I_{l+\nu}(z)\enspace,
\label{numi}
\end{equation}
where $C_l^\nu(x)$ is a Gegenbauer-polynomial,
together with \cite{EMOTa}, Chapter XI
\begin{equation}
\sum_{\mu=1}^M S_l^\mu(\Omega')S_l^\mu(\Omega'')
  ={1\over\Omega(D)}{2l+D-2\over D-2}
   C_l^{D-2\over2}(\cos\psi^{(','')})\enspace,
\end{equation}
where the $S_l^\mu(\Omega)$ are the real hyper-spherical harmonics
of degree $l$ with unit vector $\Omega$ on the $S^{(D-1)}$-sphere, $l\in
\bbbn_0$, $\Omega(D)=2\pi^{D/2}/\Gamma(D/2)$ is the volume of the
$D$-dimensional unit-sphere $S^{(D-1)}$, and $\mu=1,\dots,M$, $M=
(2l+D-2) (l+D-3)!/l!(D-3)!$. Thus for $\nu={D-3\over2}$ in (\ref{numi}),
i.e.\ on $S^{(D-2)}$
\begin{equation}
  \e^{z\cos\psi^{(','')}}=2\pi\bigg({2\pi\over z}\bigg)^{D-3\over2}
  \sum_{l=0}^\infty\sum_{\mu=1}^M
  S_l^\mu(\Omega')S_l^\mu(\Omega'')I_{l+{D-3\over2}}(z)\enspace.
\end{equation}
The angular variables in the path integral on the (D-1)-dimensional
single-sheeted hyperboloid $\CH_{-1}^{(D)}$ can be therefore separated
in a straightforward way and we obtain
\begin{eqnarray}
  & &K^{(\CH_{-1}^{(D)})}(\tau'',\tau,\{\theta''\},\{\theta'\};T)
  ={R^{1-D}\over\big(-\cosh\tau'\cosh\tau'')^{(D-2)/2}}
  \nonumber\\   & &\qquad\times
  \sum_{l=0}^\infty\sum_{\mu=1}^M S_l^\mu(\Omega')S_l^\mu(\Omega'')
  \exp\bigg[-{\i\hbar T\over8mR^2}(D-2)^2\bigg]
  K^{(\CH_{-1}^{(D)})}_l(\tau'',\tau';T)\enspace,\qquad
\label{numh}
\end{eqnarray}
and $K^{(\CH_{-1}^{(D)})}_l(\tau'',\tau';T)$ is given by
\begin{equation}
  K^{(\CH_{-1}^{(D)})}_l(\tau'',\tau';T)
  =\int\limits_{\tau(t')=\tau'}^{\tau(t'')=\tau''}\CD\tau(t)
  \exp\left\{\ih\int_{t'}^{t''}\left[{m\over2}R^2\dot\tau^2
  +{\hbar^2\over2mR^2}{(l+{D-3\over2})^2-1/4\over\cosh^2\tau}
  \right]dt\right\}\enspace.
\label{numb}
\end{equation}
Therefore we obtain similarly as before ($\tilde E=E-\hbar^2(D-2)^2/
8mR^2$)
\begin{eqnarray}
  & &\ih \int_0^{\infty}  dT\,\e^{\i TE/\hbar}
  K^{(\CH_{-1}^{(D)})}(\{\theta''\},\{\theta'\},\tau'',\tau';T)
  \nonumber\\   & &
  =\half\pi^{1-D\over2}\Gamma(\hbox{${D-1\over2}$})R^{3-D}
  \big(-\cosh\tau'\cosh\tau'')^{-(D-2)/2}
  \sum_{l=0}^\infty{2l+D-3\over D-3}C_l^{(D-3)/2}(\cos\psi^{('',')})
  \nonumber\\   & &\qquad\times
  {m\over\hbar^2}
  \Gamma\left(\sqrt{-2mR^2\tilde E}/\hbar-l-{D-4\over2}\right)
  \Gamma\left(\sqrt{-2mR^2\tilde E}/\hbar+l+{D-2\over2}\right)
  \nonumber\\   & &\qquad\times
  P_{l+(D-4)/2}^{-\sqrt{-2mR^2\tilde E}/\hbar}(\tanh\tau_<)
  P_{l+(D-4)/2}^{-\sqrt{-2mR^2\tilde E}/\hbar}(-\tanh\tau_>)\enspace.
\end{eqnarray}
Let us set $D=2d+4$ with $d=0,1,\dots$. Then $(l+{D-3\over2})^2-1/4=
(l+d)\big[(l+d)+1\big]$ and we see that in this case the radial
propagator on $\CH_{-1}^{(D)}$ yields the propagator of a reflectionless
potential \cite{CRAN}. Hence we can explicitly state for the propagator
($N=l+d$)
\begin{eqnarray}
  & &
  K^{(\CH_{-1}^{(D)})}(\{\theta''\},\{\theta'\},\tau'',\tau';T)
  \nonumber\\   & &
  ={R^{1-D}\over\big(-\cosh\tau'\cosh\tau'')^{(D-2)/2}}
  \sum_{l=0}^\infty\sum_{\mu=1}^M S_l^\mu(\Omega')S_l^\mu(\Omega'')
  \nonumber\\   & &\quad\times
  \left\{\sum_{n=0}^{N-1}\exp\left[{\i\hbar T\over2mR^2}
   \bigg((N-n)^2-{(D-2)^2\over4}\bigg)\right]\right.
  \nonumber\\   & &\quad\qquad\times
  (N-n){(2N-n)!\over n!}P_N^{n-N}(\tanh\tau')P_N^{n-N}(\tanh\tau'')
  \nonumber\\   & &\qquad\qquad
  \left.\vphantom{\sum_{n=0}^{N-1}}
  +\int_{-\infty}^\infty{dp\,p\over2\sinh\pi p}
  \exp\left[-{\i\hbar T\over2mR^2}\bigg(p^2+{(D-2)^2\over4}\bigg)\right]
  P^{-\i p}_N(\tanh\tau')P^{\i p}_N(\tanh\tau'')\right\}\qquad
\label{numa}
           \\   & &
  =\half\pi^{1-D\over2}\Gamma(\hbox{${D-1\over2}$})R^{1-D}
  \big(-\cosh\tau'\cosh\tau'')^{-(D-2)/2}
  \sum_{l=0}^\infty{2l+D-3\over D-3}C_l^{(D-3)/2}(\cos\psi^{('',')})
  \nonumber\\   & &\quad\times
  \left\{\sqrt{m\over2\pi\i\hbar T}\,\exp
   \left[-{mR^2\over2\i\hbar T}(\tau''-\tau')^2\right]
  +\half\sum_{n=0}^{N-1}\exp\left[{\i\hbar T\over2mR^2}
   \bigg((N-n)^2-{(D-2)^2\over4}\bigg)\right]\right.
  \nonumber\\   & &\qquad\qquad\times
  (N-n){(2N-n)!\over n!}
  P_N^{n-N}(\tanh\tau')P_N^{n-N}(\tanh\tau'')
  \nonumber\\   & &\qquad\qquad\times
  \left[\erf\left(\sqrt{\i\hbar T\over2mR^2}\,(N-n)
    -(\tau''-\tau')\sqrt{mR^2\over2\i\hbar T}\,\right)\right.
  \nonumber\\   & &\qquad\qquad\qquad\qquad
  \left.\left.\vphantom{\sum_{n=0}^{N-1}}
  +\erf\left(\sqrt{\i\hbar T\over2mR^2}\,(N-n)
    +(\tau''-\tau')\sqrt{mR^2\over2\i\hbar T}\,\right)\right]
  \right\}\enspace,
\end{eqnarray}
where in the resummation use has been made of the integral
representations \cite{GRA}, p.497, $\Re\beta,\Re\gamma>0$, $a>0$:
\begin{eqnarray}
  & &
  \int_0^\infty{x\,dx\over\gamma^2+x^2}\e^{-\beta x^2}\sin ax
  \nonumber\\   & &\qquad
  =-{\pi\over4}\e^{\beta\gamma^2}\bigg[2\sinh a\gamma
  +\e^{-a\gamma}
         \erf\bigg(\gamma\sqrt{\beta}\,-{a\over2\sqrt{\beta}}\bigg)
  -\e^{a\gamma}
     \erf\bigg(\gamma\sqrt{\beta}\,+{a\over2\sqrt{\beta}}\bigg)\bigg]
  \enspace,
  \nonumber\\   & &
  \int_0^\infty{dx\over\gamma^2+x^2}\e^{-\beta x^2}\cos ax
  \nonumber\\   & &\qquad
  ={\pi\over4\gamma}\e^{\beta\gamma^2}\bigg[2\cosh a\gamma
  -\e^{-a\gamma}
         \erf\bigg(\gamma\sqrt{\beta}\,-{a\over2\sqrt{\beta}}\bigg)
  -\e^{a\gamma}
      \erf\bigg(\gamma\sqrt{\beta}\,+{a\over2\sqrt{\beta}}\bigg)\bigg]
  \enspace.
  \nonumber
\end{eqnarray}
For the radial Green's function, respectively, I obtain
\begin{eqnarray}
  & &\ih \int_0^{\infty}  dT\,\e^{\i TE/\hbar}
  K^{(\CH_{-1}^{(D)})}_l(\tau'',\tau';T)
  ={1\over\hbar}\sqrt{-{m\over2E}}\,
   \exp\bigg(-|\tau''-\tau'|{\sqrt{-2mR^2E}\over\hbar}\bigg)
  \nonumber\\   & &\quad
  +\half\sum_{n=0}^{N-1}{(N-n)(2N-n)!\over-\hbar^2(N-n)^2/2mR^2-E}
   {1\over n!}P_N^{n-N}(\tanh\tau')P_N^{n-N}(\tanh\tau'')
  \nonumber\\   & &\quad\qquad\times
  \left\{1-\left(1-{\hbar(N-n)\over\sqrt{-2mR^2E}}\right)
  \cosh\left[|\tau''-\tau'|\left(N-n+{\sqrt{-2mR^2E}\over\hbar}
  \right)\right]\right\}\enspace.\qquad
\end{eqnarray}
Here use has been made of the Laplace-Fourier transformations
\cite{EMOTb}, p.177:
\begin{equation}
 \int_0^\infty dt\,\e^{(a-p)t}
      \erfc\left(\sqrt{at}+\half\sqrt{\beta\over t}\,\right)
 ={\exp(-\sqrt{a\beta}
      -\sqrt{p\beta}\,)\over\sqrt{p}\,(\sqrt{p}\,+\sqrt{a}\,)}\enspace.
  \nonumber
\end{equation}
Equation (\ref{numa}) represents the spectral expansion, where
wave-functions and energy spectra can be read off.
\goodbreak


\vglue 0.6truecm
\noindent{\bf 3.\ The ``Kepler-Problem''.}
\vglue 0.4truecm
\noindent
In the path integral (\ref{numb}) the following potential on
$\CH_{-1}^{(D)}$ is easily incorporated:
\begin{eqnarray}
  V(\tau)&=&-{q^2\over r}\sqrt{\bigg({r\over R}\bigg)^2-1}
\label{numc}
           \\   &=&
  -{q^2\over R}\tanh\tau\enspace,
\label{numd}
\end{eqnarray}
where $r^2=\sum_{i=1}^{D-1}x_i^2\geq R^2$. For $D=4$ (\ref{numc}) has
the structure of a Kepler-problem in a space of constant curvature
\cite{BIJ, GROg}. In our case of the single-sheeted hyperboloid we want
to keep this notion for every dimension $D$, and we will see that a
similar structure familiar from the usual Coulomb-problem in the energy
spectrum will in fact arise, however with some significant different
features. Furthermore, the potential (\ref{numd}) is not singular for
any value of $\tau\in\bbbr$.

Implementing the potential (\ref{numd}) in the radial path integral
(\ref{numb}) yields
\begin{eqnarray}
  & &\!\!\!\!
  K^{(q^2)}_l(\tau'',\tau';T)
  \nonumber\\   & &\!\!\!\!
  =\int\limits_{\tau(t')=\tau'}^{\tau(t'')=\tau''}\CD\tau(t)
  \exp\left\{\ih\int_{t'}^{t''}\left[{m\over2}R^2\dot\tau^2
  +{\hbar^2\over2mR^2}{(l+{D-3\over2})^2-1/4\over\cosh^2\tau}
  +{q^2\over R}\tanh\tau\right]dt\right\}
  \enspace.\qquad
\label{numg}
\end{eqnarray}
Equation (\ref{numg}) has the form of the path integral for the
Rosen-Morse potential
\begin{equation}
  V(x)=-{B\over\cosh^2x/R}+A\tanh{x\over R}\enspace,
\end{equation}
($A,B,R$ constants, $x\in\bbbr$) which has been discussed in
\cite{GROe, KLEMUS} by means of the path integral of the modified
P\"oschl-Teller potential, c.f.\ the Appendix. Identifying
\begin{equation}
  A=-{q^2\over R},\qquad
  B=\hbar^2{(l+{D-2\over2})^2-\viert\over2mR^2},\qquad\tau={x\over R}
  \enspace,
\end{equation}
gives the path integral solution
\begin{eqnarray}
 & &\!\!\!\!\!\!\!\!
  \ih\int_0^\infty  dT\,\e^{\i TE/\hbar}\!\!
  \int\limits_{\tau(t')=\tau'}^{\tau(t'')=\tau''}\!\!\CD\tau(t)
  \exp\left\{\ih\int_{t'}^{t''}\left[{m\over2}R^2\dot\tau^2
  +\hbar^2{(l+{D-3\over2})^2-\viert\over2mR^2\cosh^2\tau}
  +{q^2\over R}\tanh\tau\right]dt\right\}
  \nonumber\\   & &\!\!\!\!\!\!\!\!
  ={mR^2\over\hbar^2}{\Gamma(m_1-L_B)\Gamma(L_B+m_1+1)\over
           \Gamma(m_1+m_2+1)\Gamma(m_1-m_2+1)}
  \nonumber\\   & &\times
  \bigg({1-\tanh\tau'\over2}\cdot
                            {1-\tanh\tau''\over2}\bigg)^{m_1-m_2\over2}
  \bigg({1+\tanh\tau'\over2}\cdot
                            {1+\tanh\tau''\over2}\bigg)^{m_1+m_2\over2}
  \nonumber\\   & &\times
  {_2}F_1\bigg(-L_B+m_1,L_B+m_1+1;m_1+m_2+1;{1+\tanh\tau_>\over2}\bigg)
  \nonumber\\   & &\times
  {_2}F_1\bigg(-L_B+m_1,L_B+m_1+1;m_1-m_2+1;{1-\tanh\tau_<\over2}\bigg)
           \\   & &\!\!\!\!\!\!\!\!
  =\sum_{n=0}^{N_M}{\Psi^{(q^2)\,*}_{n,l}(\tau')
                    \Psi^{(q^2)}_{n,l}(\tau'')\over E_{n,l}^{(q^2)}-E}
  +\int_0^\infty dp{\Psi^{(q^2)\,*}_{p,l}(\tau')
   \Psi^{(q^2)}_{p,l}(\tau'')\over E_{p,l}^{(q^2)}-E}\enspace.
  \end{eqnarray}
Here are $L_B=l+{D-4\over2}$, $m_{1,2}=\sqrt{m/2}\,R\Big(\sqrt{-q^2/R
-E} \pm\sqrt{q^2/R-E}\Big)/\hbar$, and $\tau_{<,>}$ the smaller/ larger
of $\tau',\tau''$, respectively. The wave-functions and the
energy-spectrum are given by \Big[$s\equiv2l+D-3$, $n=0,\dots,N_M<l
+{D-4\over2} -\sqrt{R/a}$ with $a=\hbar^2/mq^2$ the Bohr radius,
$k_1=\half(1+s)$, $k_2=\half\big[1+\half(s-2n-1)-{2mq^2R\over\hbar(s-2n-
1)}\big]$, $u=\half(1+\tanh\tau)$, note $k_2-\half>0$\Big]:
\begin{eqnarray}
  \Psi^{(q^2)}_{n,l}(\tau)
                &=&
  \bigg[\bigg({1\over R}+{4mq^2\over\hbar^2(s-2n-1)^2}\bigg)
   {(s-2k_2-2n)n!\,\Gamma(s-n)\over
   \Gamma(s+1-n-2k_2)\Gamma(2k_2+n)}\bigg]^{1/2} 2^{n+(1-s)/2}
  \nonumber\\   & &\qquad\vphantom{\bigg[}\times
     (1-\tanh\tau)^{\half s-k_2-n}
     (1+\tanh\tau)^{k_2-\half}P_n^{(s-2k_2-2n,2k_2-1)}(\tanh\tau)
  \enspace,\qquad
  \\
  E_{n,l}^{(q^2)}
  &=&-\left[\hbar^2{(l+{D-4\over2}-n)^2\over2mR^2}
          +{mq^4\over2\hbar^2(n-l-{D-4\over2})^2}\right]\enspace.
\label{nume}
  \end{eqnarray}
The wave-functions and the energy-spectrum of the continuous states are
given by $\Big[k_2\equiv\half(1+\i\tilde p)$  $\kappa=\half(1+\i p)$,
$\tilde p\equiv\sqrt{2mR^2(-2q^2/R+\hbar^2p^2/2mR^2)}/\hbar>0\Big]$:
\begin{eqnarray}
  \Psi_{p,l}^{(q^2)}(x)&=&N_p^{(k_1,k_2)}
  (1-u)^{-\i p/2}u^{\i\tilde p/2}
  \nonumber\\   & &\qquad\qquad\vphantom{\bigg[}\times
  {_2}F_1\{\bhalf[1+s+\i(\tilde p-p)],
  \bhalf[1-s+\i(\tilde p-p)];1+\i\tilde p;u\}\enspace,
           \\
  N_k^{(k_1,k_2)}&=&{1\over\Gamma(2k_2)}\sqrt{p\sinh\pi p\over2\pi^2}
  \Big[\Gamma(k_1+k_2-\kappa)\Gamma(-k_1+k_2+\kappa)
  \nonumber\\   & &\qquad\qquad\vphantom{\bigg[}\times
  \Gamma(k_1+k_2+\kappa-1)\Gamma(-k_1+k_2-\kappa+1)\Big]^{1/2}\enspace,
           \\
  E_{p,l}&=&{\hbar^2p^2\over2mR^2}-{q^2\over R}\enspace.
  \end{eqnarray}
In the limit $q^2=0$ the case of (\ref{numa}) is easily recovered.
Note that for the entire problem the additional ``zero-energy'' shift
$E_0^{(D)}=\hbar^2(D-2)^2/8mR^2$ has to be taken into account, c.f.\
(\ref{numh}). We see that the energy spectrum (\ref{nume}) of the
``Kepler-problem'' on the single-sheeted hyperboloid has in fact a form
familar from the usual Coulomb-problem in flat space, and in spaces of
(positive and negative) constant curvature \cite{BIJ, GROg},
respectively. However, in the present example the flat space limit
($R\to\infty$) does not make any sense, and the corresponding Hilbert
space does not exist.
\goodbreak


\vglue 0.6truecm
\noindent{\bf 4.\ The Constant Magnetic Field.}
\vglue 0.4truecm
\noindent
Let us introduce on $\CH_{-1}^{(3)}$ the vector-potential $\vec A$
\begin{equation}
  \vec A=(A_\tau,A_\phi)=\i B\sinh\tau(0,1)\enspace.
\end{equation}
The magnetic field is thus calculated to read as $dB=(\partial_\tau
A_\phi-\partial_\phi A_\tau)d\tau\wedge d\phi =\i B\cosh\tau d\tau
\wedge d\phi = B\,\sqrt{\det(g_{ab})}$ which has the form $constant
\times volume$-$form$ and can therefore be interpreted as a constant
field on $\CH_{-1}^{(3)}$. Note the imaginary unit involved in $\vec A$
which is due to the indefinite metric of $\CH_{-1}^{(D)}$. The path
integral with the vector potential $\vec A$ then has the form
($b=eB/\hbar c$)
\begin{eqnarray}
  & &
  K^{(\CH_{-1}^{(3)},b)}(\tau'',\tau',\phi'',\phi';T)
  \nonumber\\   & &
  ={1\over R^2}
   \lim_{N\to\infty}\bigg({mR^2\over 2\pi\epsilon\hbar}\bigg)^{N}
  \prod_{j=1}^{N-1}\int_{-\infty}^\infty\cosh\tau_jd\tau_j
  \int_0^{2\pi}d\phi_j
  \nonumber\\   & &\qquad\times
  \exp\left\{\ih\sum_{j=1}^{N}\left[{mR^2\over2\epsilon}\Big(
   \Delta^2\tau_j-{\widehat{\cosh}}^2\tau_j\Delta^2\phi_j\Big)
  -\i\hbar b\,{\widehat{\sinh}}\tau_j\Delta\phi_j
  -{\epsilon\hbar^2\over8mR^2}\bigg(1+{1\over\cosh^2\tau_j}\bigg)
   \right]\right\}
  \nonumber\\   & &
  \\            & &
   \equiv{1\over R^2}
   \int\limits_{\tau(t')=\tau'}^{\tau(t'')=\tau''}\cosh\tau\CD\tau(t)
   \int\limits_{\phi(t'')=\phi''}^{\phi(t'')=\phi''}\CD\phi(t)
  \nonumber\\   & &\qquad\times
   \exp\left\{\ih\int_{t'}^{t''}\left[{m\over2}R^2\Big(
   \dot\tau^2-\cosh^2\tau\dot\phi^2\Big)
   -\i\hbar b\sinh\tau\dot\phi
   -{\hbar^2\over8mR^2}\bigg(1+{1\over\cosh^2\tau}\bigg)
  \right]dt\right\}\enspace.\qquad
\end{eqnarray}
We perform a Fourier expansion according to
\begin{eqnarray}
  & &K^{(\CH_{-1}^{(3)},b)}(\tau'',\tau',\phi'',\phi';T)
  ={1\over2\pi R^2}\Big(-\cosh\tau'\cosh\tau''\Big)^{-1/2}
  \nonumber\\   & &\qquad\qquad\qquad\times
  \exp\bigg[-{\i\hbar T\over2mR^2}\bigg(b^2+\viert\bigg)\bigg]
  \sum_{l=-\infty}^\infty\e^{\i l(\phi''-\phi')}
  K^{(\CH_{-1}^{(3)},b)}_l(\tau'',\tau';T)\enspace,
  \\   & &
  K^{(\CH_{-1}^{(3)},b)}_l(\tau'',\tau';T)
  ={1\over2\pi}\int_0^{2\pi}d\phi''\,\e^{-\i l(\phi''-\phi')}
  K^{(\CH_{-1}^{(3)},b)}(\tau'',\tau',\phi'',\phi';T)\enspace.
\end{eqnarray}
We therefore obtain that the radial kernel
$K^{(\CH_{-1}^{(3)},b)}_l(T)$ is given by
\begin{eqnarray}
  & &
  K^{(\CH_{-1}^{(3)},b)}_l(\tau'',\tau';T)
  \nonumber\\   & &
  =\int\limits_{\tau(t')=\tau'}^{\tau(t'')=\tau''}\CD\tau(t)
  \exp\left\{\ih\int_{t'}^{t''}\left[{m\over2}R^2\dot\tau^2
  +{\hbar^2\over2mR^2}\bigg({l^2+b^2-1/4\over\cosh^2\tau}
      +2\i lb{\tanh\tau\over\cosh\tau}\bigg)\right]dt\right\}
  \enspace,\qquad
\end{eqnarray}
which is the path integral of a barrier tunneling potential $V(x)=
(\hbar^2/2m)(A+B\tanh x/\cosh x+C\tanh^2x)$ as discussed in \cite{GROs}
(compare \cite{PERT} for a detailed study of reflection and scattering
properties) and belongs to a class of potentials called Scarf-like
potentials \cite{DKS}. We perform the coordinate transformation
$(1+\i\sinh\tau)/2=\cosh^2z$ and obtain ($M=4m$)
\begin{eqnarray}
  K^{(\CH_{-1}^{(3)},b)}_l(\tau'',\tau';T)
  &\!\!\to&\!\!\int\limits_{z(t')=z'}^{z(t'')=z''}\CD z(t)
  \exp\left\{\ih\int_{t'}^{t''}\bigg[{M\over2}\dot z^2
  -{\hbar^2\over2M}\bigg({C-\i B\over\sinh^2z}
  -{C+\i B\over\cosh^2z}\bigg)\bigg]ds\right\}
  \nonumber\\
  &\!\!=&\!\!
  \half\Bigg\{\sum_{n=0}^{N_M}\e^{-\i TE_n^{(\CH_{-1}^{(3)},b)}/\hbar}
  \Psi_n^{(\CH_{-1}^{(D)},b)*}(z') \Psi_n^{(\CH_{-1}^{(3)},b)}(z'')
  \nonumber\\   & &\qquad\qquad\qquad
  +\int_{-\infty}^\infty dp\,\e^{-\i TE_p^{(\CH_{-1}^{(3)},b)}/\hbar}
  \Psi_p^{(\CH_{-1}^{(3)},b)*}(z')
  \Psi_p^{(\CH_{-1}^{(3)},b)}(z'')\Bigg\}\enspace,\qquad
\end{eqnarray}
We do not worry about the fact that this is a complex coordinate
transformation (compare also \cite{BIJ} in the treatment of the Kepler
problem in a space of constant positive curvature, where an even more
complicated coordinate transformation has been made, accompanied by an
additional time-transformation). Due to the specific nature of the
vector potential we have chosen, the latter path integral is a usual
one-dimensional path integral with a {\it real\/} potential [e.g.\ $C\pm
\i B=(l\pm b)^2-\viert$]. Here $k_1=\half(1+\sqrt{C+\i B+1/4}\,)=\half(1
+|l+b|)$, $k_2=\half(1+|l-b|)$ in the notation of the Appendix (the
correct signs of the square roots follow from the vanishing of the
bound state wave-functions for $x\to\pm\infty$). Therefore we obtain
for the energy spectrum for the motion $\CH_{-1}^{(3)}$ with a constant
magnetic field
\begin{equation}
  E_n^{(\CH_{-1}^{(3)},b)}={\hbar^2\over2mR^2}\bigg[\viert+b^2
   -(n+\bhalf-\bhalf|l+b|+\bhalf|l-b|)^2\bigg]\enspace.
\end{equation}
We have $n=0,1,2,\dots,N_M<\half(|l+b|-|l-b|-1)$. For the bound state
wave functions we get (reinserting $z\to\tau$)
\begin{eqnarray}
  \Psi_{n,l}^{(\CH_{-1}^{(3)},b)}(\tau)
  &=&\bigg[{(|l+b|-|l-b|-n-1)n!\Gamma(|l+b|-n)\over
         \Gamma(|l+b|-|l-b|-n)\Gamma(|l-b|+n+1)}\bigg]^\half
  \nonumber\\   & &\times
  \bigg({\i\sinh\tau-1\over 2}\bigg)^{\half(\half+|l-b|)}
  \bigg({\i\sinh\tau+1\over 2}\bigg)^{\half(\half-|l+b|)}
  P_n^{(|l-b|,-|l+b|)}(\i\sinh\tau)\enspace.\qquad
\end{eqnarray}
and the $P_n^{(a,b)}(z)$, $z\in\bbbc$, are Jacobi polynomials.

Rescaling the parameter $p$ according to $p\to 2p$ in the $p$-integral
for the continuous states we get for the continuous spectrum
\begin{equation}
  E_p^{(\CH_{-1}^{(D)},b)}={\hbar^2\over2mR^2}\bigg(p^2+b^2+\viert\bigg)
  \enspace,
\end{equation}
and the wave functions have the form
\begin{eqnarray}
  \Psi_{p,l}^{(\CH_{-1}^{(3)},b)}(\tau)
  &=&{\sqrt{p\sinh2\pi p}\over\pi\Gamma(1+|l-b|)}
  \Big|\Gamma\big[\bhalf(1+|l-b|+|l+b|)-\i p\big]
       \Gamma\big[\bhalf(1+|l-b|-|l+b|)-\i p\big]\Big|
  \nonumber\\   & &\times
  \bigg({\i\sinh\tau+1\over2}\bigg)^{\half(\half+|l-b|)}
  \bigg({\i\sinh\tau-1\over2}\bigg)^{\i p-\half(\half+|l-b|)}
  \nonumber\\   & &\times
  {_2}F_1\bigg[\half\big(1+|l-b|+|l+b|\big)-\i p,
  \nonumber\\   & &\qquad\qquad\qquad
               \half\big(1+|l-b|-|l+b|\big)-\i p;1+|l-b|;
               {\i\sinh\tau-1\over\i\sinh\tau+1}\bigg]\enspace.
\end{eqnarray}
\goodbreak


\vglue 0.6truecm
\noindent{\bf 5.\ Summary and Discussion.}
\vglue 0.4truecm
\noindent
In this note I have studied path integration on the (D-1)-dimensional
single-sheeted hyperboloid in a conveniently chosen coordinate system of
$(1,D-1)$-dimensional pseudo-bispherical polar coordinates: first the
two-dimensional, second its higher-dimensional generalization, third a
potential problem, and finally the case of a constant magnetic field. In
all cases the propagators, the Green functions, and the corresponding
wave-functions and energy spectra could be easily determined by the
formalism. We found that in comparison to the (two-sheeted) pseudosphere
bound states are allowed already for the free motion on the
single-sheeted hyperboloid, where the number of bound stated is
determined by the angular momentum number. Similarly as in the case of
the pseudosphere \cite{GRSc}, the hyperbolic plane with magnetic fields
\cite{GROf}, and other hyperbolic spaces \cite{GROn}, a ``zero-energy''
shift $E_0^{(D)}=\hbar^2(D-2)^2 /8mR^2$ appeared in the energy spectra.
We also found that in even dimensions $D$ the corresponding ``radial''
propagator for the free motion has the form of a reflectionless
potential propagator, a property which allows simplifications in the
explicit form of the radial propagator.

The potential problem on $\CH_{-1}^{(D)}$ which was studied, we called
``Kepler-problem'' on $\CH_{-1}^{(D)}$ due to its general structure in
terms of the coordinates of the embedding space. The corresponding path
integral could be reduced to a known path integral solution, namely of
the path integral for the Rosen-Morse potential. However, as we saw, it
cannot be interpreted as a genuine Coulomb-problem as known from the
other spaces of constant curvature because it is not singular, it is
the solution of the homogeneous Laplace equation (and not of the
inhomogeneous one), and the flat space limit does not make sense.

In the forth Section we discussed the case of constant magnetic field
on the two-dimensional single-sheeted hyperboloid. Here another
path integral identity came into play, i.e.\ the path integral
solution from a specific form of a Scarf-like potential, respectively
a hyperbolic barrier potential.

In all our problems, free motion, ``Kepler-problem'', and the constant
magnetic field, we could observe a nice interplay between motion in
spaces of constant curvature on the one side, and potential problems
emerging from them by separating the angular variables on the other.
This feature is well-known also from other realizations of (real)
Lobachevskian spaces \cite{GROf, GROc, GROn}. It has its origin in the
underlying group structure of the space in question, respectively the
corresponding dynamical group of the potential problem \cite{INO, IKG},
where the most known example is the Hydrogen atom in flat with its
$\O(4)$ symmetry. The pseudo-bispherical coordinates coming from the
$\SO(1,D-1)$ group structure of $\CH_{-1}^{(D)}$ allow the separation
of the angular variables due to $\SO(m,n)\supset\SO(m)\times\SO(n)$,
and the remaining path radial- (i.e.\ $\tau$-) path integration can be
transformed into a $\SU(1,1)$ path integration.

{}From the present model no quantum mechanical discussion seems to have
been made until now, an operator approach as well as a path integral
approach. The solution of path integration on $\CH_{-1}^{(D)}$,
together with the potential problem and the case of a magnetic field,
has in comparison to an operator approach the advantage of presenting
a global picture of the quantum theory in question, whereas the
Schr\"odinger approach allows only a local one, and the explicit form
of the Feynman kernel gives the complete solution in terms of the
wave-functions and the energy-spectrum, respectively. The examples
demonstrate once more the consistency as well as the universal utility
and feasibility of the Feynman path integral and of our general method
developed in \cite{GRSb}.
\goodbreak

\vglue 0.6truecm
\noindent{\bf Acknowledgement.}
\vglue 0.4truecm
\noindent
I would like to thank E.Rabinovici for drawing my attention to the
problem of the single-sheeted hyperboloid.
\goodbreak


\vglue 0.6truecm
\noindent{\bf Appendix.}
\vglue 0.4truecm
\noindent
In this Appendix we cite an important path integral identity important
for the discussion in the text. Let us consider quantum mechanical
models related to the modified P\"oschl-Teller (mPT) potential
\begin{equation}
  V^{(mPT)}(\tau)={\hbar^2\over2m}\left({\eta^2-\viert\over\sinh^2\tau}
    -{\nu^2-\viert\over\cosh^2\tau}\right),\qquad\tau>0\enspace,
\end{equation}
which has a (hidden) $\SU(1,1)$ symmetry. The path integral solution is
due to \cite{BJb, INOWI} and has the form (we use the notation of Ref.\
\cite{FW} for the bound and continuous states, respectively,
$k_1=\half(1\pm\nu)$, $k_2=\half(1\pm\eta)$, for the explicit form of
the Green function compare \cite{GROs, KLEMUS})
\begin{eqnarray}
  & &\ih \int_0^\infty dT\,\e^{\i TE/\hbar}\!\!\!
  \int\limits_{r(t')=r'}^{r(t'')=r''}\!\!\!\CD r(t)
  \exp\left\{\ih\int_{t'}^{t''}\left[{m\over2}\dot r^2
   -{\hbar^2\over2m}\bigg({\eta^2-\viert\over\sinh^2r}
   -{\nu^2-\viert\over\cosh^2r}\bigg)\right]dt\right\}
  \nonumber\\   & &
  ={m\over\hbar^2}{\Gamma(m_1-L_\nu)\Gamma(L_\nu+m_1+1)\over
   \Gamma(m_1+m_2+1)\Gamma(m_1-m_2+1)}
  \nonumber\\   & &\qquad\times\vphantom{\Big(}
  \big(\cosh r'\cosh r''\big)^{-(m_1-m_2)}
  \big(\tanh r'\tanh r''\big)^{m_1+m_2+1/2}
  \nonumber\\   & &\qquad\times
  {_2}F_1\bigg(-L_\nu+m_1,L_\nu+m_1+1;m_1-m_2+1;
                          {1\over\cosh^2r_<}\bigg)
  \nonumber\\   & &\qquad\times
  {_2}F_1\bigg(-L_\nu+m_1,L_\nu+m_1+1;m_1+m_2+1;\tanh^2r_>\bigg)
  \\   & &
  =\sum_{n=0}^{N_M}
  {\Psi^{(k_1,k_2)\,*}_n(r')\Psi^{(k_1,k_2)}_n(r'')\over E_n-E}
  +\int_0^\infty dp{\Psi^{(k_1,k_2)\,*}_p(r')\Psi^{(k_1,k_2)}_p(r'')
   \over \hbar^2p^2/2m-E}\enspace,
\end{eqnarray}
($m_{1,2}=\half(\eta\pm\sqrt{-2mE}/\hbar$, $L_\nu=\half(1-\nu$).
The correct signs depend on the boundary conditions for $r\to0$ and
$r\to\infty$, respectively. Here we have introduced the Green function
\begin{equation}
  G(q'',q';E)
  :=\ih\int_0^\infty dT\,\e^{\i(E+\i\epsilon)T/\hbar}K(q'',q';T)
  =\bigg<q''\bigg|{1\over H-E-\i\epsilon}\bigg|q'\bigg>\enspace,
\end{equation}
($\b H$ the Hamiltonian) where a small positive imaginary part
$(\epsilon>0)$ has been added to the energy $E$. (We shall not
explicitly write the $\i\epsilon$, but will tacitly assume that the
various expressions are regularized according to this rule). The bound
state wave-functions are e.g.\ given by $(\kappa=k_1-k_2-n$)
\begin{eqnarray}
  \Psi_n^{(k_1,k_2)}(\tau)
  &=&
  {1\over\Gamma(2k_2)}\bigg[{2(2\kappa-1)
     \Gamma(k_1+k_2-\kappa)\Gamma(k_1+k_2+\kappa-1)
    \over\Gamma(k_1-k_2+\kappa)\Gamma(k_1-k_2-\kappa+1)}\bigg]^{1/2}
  \nonumber\\   & &\qquad\qquad\times\vphantom{\Big(}
  (\sinh\tau)^{2k_2-\half}(\cosh\tau)^{-2k_1+3/2}
  \nonumber\\   & &\qquad\qquad\times\vphantom{\Big(}
  {_2}F_1(-k_1+k_2+\kappa,-k_1+k_2-\kappa+1;2k_2;-\sinh^2\tau)\enspace,
           \\
  E_n&=&
  -{\hbar^2\over2m}\bigg[2(k_1-k_2-n)-1\bigg]^2
  \enspace.\qquad
\end{eqnarray}
Here is $n=0,1,\dots,N_m<k_1-k_2-1/2$, with $N_M$ the maximum number
of bound states. The continuous states have the form $[\kappa=
\half(1+\i p)]$
\begin{eqnarray}
  \Psi_p^{(k_1,k_2)}(\tau)
  &=&
  N_p^{(k_1,k_2)}(\cosh\tau)^{\kappa-2k_2-1/2}(\sinh\tau)^{2k_2-1/2}
  \nonumber\\   & &\qquad\qquad\qquad\times\vphantom{\Big(}
    {_2}F_1(k_1+k_2-\kappa,k_2-k_1-\kappa+1;2k_2;\tanh^2\tau)\enspace,
  \\
  N_p^{(k_1,k_2)}
  &=&
  {1\over\Gamma(2k_2)}\sqrt{p\sinh\pi p\over2\pi^2}
    \Big[\Gamma(k_1+k_2-\kappa)\Gamma(-k_1+k_2+\kappa)
  \nonumber\\   & &\qquad\qquad\qquad\times
    \Gamma(k_1+k_2+\kappa-1)\Gamma(-k_1+k_2-\kappa+1)\Big]^{1/2}
  \enspace,
\end{eqnarray}
and $E_p=\hbar^2p^2/2m$. Here the functions ${_2}F_1(a,b;c;z)$
($z\in\bbbc$) denote hypergeometric functions.

Of particular importance is the following special case, where a path
integral solution according to \cite{ANAN, GROc, KLEMUS} has the form
($n=0,1,\dots,N_M<l-\half$, $l>0$, $x\in\bbbr$)
\begin{eqnarray}
  & &\ih \int_0^\infty dT\,\e^{\i TE/\hbar}
  \int\limits_{x(t')=x'}^{x(t'')=x''}\CD x(t)
  \exp\left[\ih\int_{t'}^{t''}\bigg({m\over2}\dot x^2
  +{\hbar^2\over2m}{l^2-{1\over4}\over\cosh^2x}\bigg)dt\right]
  \nonumber\\   & &\qquad
  ={m\over\hbar^2}
  \Gamma\bigg({1\over\hbar}\sqrt{-2mE}-l+\half\bigg)
  \Gamma\bigg({1\over\hbar}\sqrt{-2mE}+l+\half\bigg)
  \nonumber\\   & &\qquad\qquad\times\vphantom{\Big(}
  P_{l-1/2}^{-\sqrt{-2mE}/\hbar}(\tanh x_<)
  P_{l-1/2}^{-\sqrt{-2mE}/\hbar}(-\tanh x_>)
           \\   & &\qquad
  =\sum_{n=0}^{N_M}\bigg(n-l-\half\bigg){\Gamma(2l-n)\over n!}
  \frac{P_{l-1/2}^{n-l+\half}(\tanh x')P_{l-1/2}^{n-l+\half}(\tanh x'')}
  {-\hbar^2(n-l+\half)^2/2m-E}
  \nonumber\\   & &\qquad\qquad\qquad
  +\half\int_{-\infty}^{\infty}\,{dp\,p\sinh\pi p\over\hbar^2p^2/2m-E}
    {P^{\i p}_{l-1/2}(\tanh x')P^{-\i p}_{l-1/2}(\tanh x'')
     \over\cos^2\pi l+\sinh^2\pi p}\enspace.
\end{eqnarray}
Here $P_\nu^\mu(x)$ are Legendre functions of the first kind.
\goodbreak


\end{document}